\documentclass{JINST}

\usepackage{amssymb}
\usepackage{lineno}
\usepackage{graphicx}
\usepackage{amssymb}
\usepackage{amsmath}
\usepackage{epstopdf}
\usepackage{relsize}

\def\cesrta{{C{\smaller[2]ESR}TA}}

\def\cesrta{{C{\smaller[2]ESR}TA}}

\newcommand{\beq} {\begin{equation}}
\newcommand{\eeq} {\end{equation}}

\title{The Conversion of CESR to Operate as the Test Accelerator, CesrTA, Part 1: Overview}

\author{M.G.Billing\\
Cornell University, \\
Ithaca, NY}

\abstract{ Cornell's electron/positron storage ring (CESR) was
modified over a series of accelerator shutdowns beginning in May
2008, which substantially improves its capability for research and
development for particle accelerators.  CESR's energy span from 1.8
to 5.6 GeV with both electrons and positrons makes it 
ideal for the study of a wide spectrum of
accelerator physics issues and instrumentation related to present
light sources and future lepton damping rings. Additionally a number
of these are also relevant for the beam physics of proton
accelerators. This paper outlines the motivation, design and
conversion of CESR to a test accelerator, {\cesrta}, enhanced to
study such subjects as low emittance tuning methods, electron cloud
(EC) effects, intra-beam scattering, fast ion instabilities as well
as general improvements to beam instrumentation.  While the initial
studies of {\cesrta} focussed on questions related to the
International Linear Collider (ILC) damping ring design, {\cesrta}
is a very flexible storage ring, capable of studying a wide range of
accelerator physics and instrumentation questions.  This paper
contains the outline and the basis for a set of papers documenting
the reconfiguration of the storage ring and the associated
instrumentation required for the studies described above.  Further
details may be found in these papers.}


\begin{document}

\section{CesrTA Collaboration}

J.P.~Alexander$^{1}$, F.~Antoniou$^{9}$, D.~Asner$^{7}$,
R.P.~Badman$^{26}$, J.~Barley$^{1}$, L.~Bartnik$^{1}$,
M.G.~Billing$^{1}$, L.~Boon$^{24}$, K.R.~Butler$^{1}$,
J.~Byrd$^{21}$
% ...
\\
\llap{$^{30}$}University of Utah,\\
Department of Physics and Astronomy, Salt Lake City, UT 84112,
U.S.A.


\section{Overview of \cesrta\ Program and CESR Modifications}
\label{sec:cesr_conversion.overview}

The \cesrta\ program was initially proposed to study accelerator physics questions related to the design of 
the electron and positron damping rings for the ILC accelerator complex.  Several projects were determined
to be appropriate for the CESR storage ring, which operates in the same ring with either electrons or positrons
at energies comparable to those proposed for the damping rings.  These goals group into five distinct categories with the associated references detailing some of the recent results of these studies:

\begin{itemize}
\item Developing techniques for tuning beams in storage rings for low vertical emittances.\cite{PRSTAB17:044003}
\item Studying the dynamics for positron bunches within trains of bunches due to the growth of electron clouds along the trains.\cite{IPAC14:TUPRI067}
\item Creating strategies to mitigate the effects of electron clouds on the positron bunches within trains.\cite{NIMA770:141to154, ARXIV:1309.2625, NIMA760:86to97}
\item Characterizing the effects of intra-beam scattering on single bunches.\cite{PRSTAB16:104401,PRSTAB17:044002,IPAC14:TUPRI035}
\item Studying the effects of fast ion instabilities on the electron bunches within trains.\cite{SUBMITTED:FII1,IPAC14:TUPRI036}
\end{itemize}

An additional overall goal for the \cesrta\ program is to record and archive the data that is being acquired in order to make this available to accelerator physicists for testing future models of the phenomena that were studied as part of the program.  As a consequence of this, the documentation of the storage ring's configuration, the accelerator environment in the experimental regions and the operation of the diagnostic instrumentation needs to be sufficiently complete for those, who would utilize this data in the future.

The conversion of CESR to permit the execution of the \cesrta\ program required several extensive modifications.  These included significant reconfiguring of CESR's accelerator optics by removing the CLEO high energy physics detector and its interaction region, moving six superconducting wigglers and reconfiguring the L3 straight section.  There were also major vacuum system modifications to accommodate the changes in layout of the storage ring guide-field elements, to add electron cloud diagnostics and to prepare regions of the storage ring to accept beam pipes for the direct study of electron clouds (EC).  A large variety of instrumentation was also developed to support new electron cloud diagnostics, to increase the capabilities of the beam stabilizing feedback systems and the beam position monitoring system, to develop new X-ray beam size diagnostics and to increase the ability for studying beam instabilities.  The entire effort to reconfigure CESR and to install the new instrumentation took place over four separate accelerator down periods. This conversion process and the instruments developed for the \cesrta\ program are described broadly in the following sections.  These include the \cesrta\ optics design, the accelerator lattice modifications and a general overview of instrumentation and controls upgrades.  

This is the first of a four part series describing the conversion of CESR for the \cesrta\ program.  This paper describes an overview of the conversion project, including details on the accelerator lattice, magnets and controls, alignment and survey, injection controls and a broad overview of beam instrumentation.  Part two documents the vacuum system modifications for the project.  Required beam instrumentation upgrades are presented in part three.  Part four describes the modification of superconducting wigglers to accommodate vacuum chambers containing retarding field analyzers for electron cloud spatial measurements and EC mitigations.  More detailed discussion of the accelerator modifications and instrumentation may be found in the subsequent papers.

The initial phases of the \cesrta\ program benefitted from the collaborative efforts of accelerator physicists and engineers from many laboratories throughout the world.  As this paper provides an overview of the 
initial \cesrta\ 
program and modification of the storage ring CESR to permit these studies, it seems appropriate to list
our collaborators as co-authors to this paper to acknowledge our appreciation for their thoughts, suggestions
and efforts.  


\section {{\cesrta} Lattice and
Layout} \label{sec:introduction.parameters_reach}

The CESR storage ring, outfitted with independently powered
quadrupoles and sextupoles and 18~meters of superconducting damping
wigglers is capable of operating with a very wide range of storage ring
optics. Basic parameters for the storage ring are found in
Table \ref{tab:intro:parameters}.  The LINAC and synchrotron provide
full energy injection of electron and positron beams. The storage ring
magnets allow operation over an energy range of 1.8 to
5.6~GeV. Multibunch feedback systems stabilize motion in all three
planes for trains of bunches with bunch spacing of as short as 4~ns. The four
single cell 500~MHz superconducting RF cavities operate over a range of
gradients that enables investigations of bunch length dependencies.
The complement of more than 100 steering correctors and 25 skew quad
correctors is sufficient to
eliminate relevant sources of vertical emittance dilution.

\begin{table}[htb]
\begin{center}
\caption[Basic CESR Storage Ring Parameters]{\label{tab:intro:parameters}Basic CESR Parameters} \vspace*{1ex}
\begin{tabular}{|l|c|c|c|c|c|}
\hline
Parameter & Value & Units\\
\hline
Ring Circumference & 768.438 & m \\
Circulation Period & 2.56 & $\mu$sec \\
Operating Energy Range & 1.8 - 5.6 & GeV\\
RF System Frequency & 499.759 & MHz \\
Number of RF Buckets & 1281 & \\
Species & Electrons \& &\\
  & Positrons & \\
Maximum Single Beam Current & 250 & mA\\
  \hline
Any Combination of Bunches for & & \\
  183 of 183 Buckets at a Spacing of & 14 & nsec\\
  \hline
Any Combination of Bunches for &  & \\
  First 600 Buckets at a Spacing of & 4 & nsec\\
  \hline
\end{tabular}
\end{center}
\end{table}

\subsection{Low Emittance Optics}
\begin{figure}[tb] 
   \centering
   \includegraphics[width=4in]{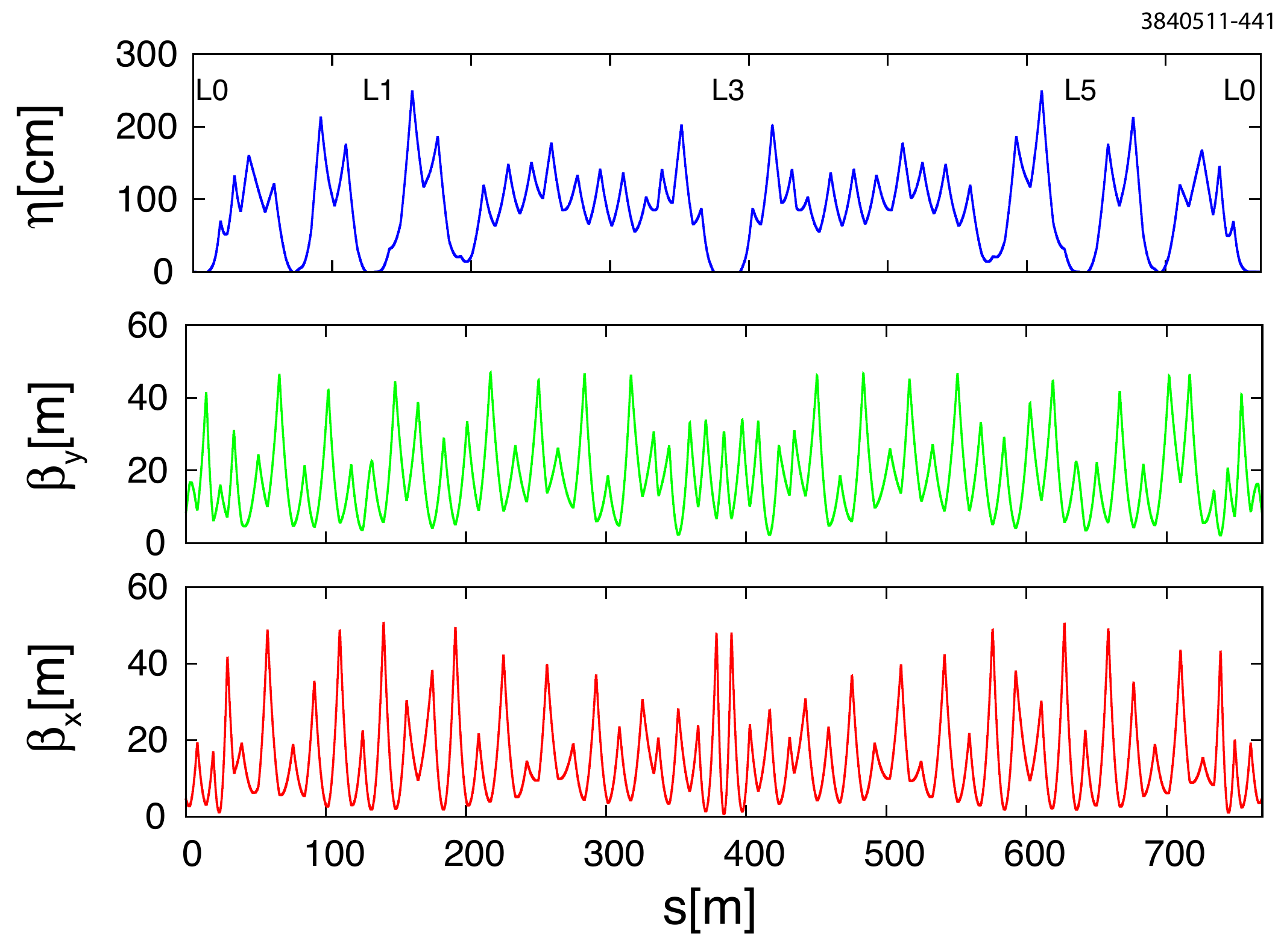}
  \caption[Optics functions of low emittance ($\epsilon_x=2.6$
    nm-rad) lattice.]{\label{fig:intro:magnetic_optics} Design optics functions of low emittance ($\epsilon_x=2.6$
    nm-rad) lattice, with horizontal dispersion, $\eta$, at the top and the vertical and
    horizontal amplitude functions, $\beta_y$ and $\beta_x$ below.  Regions of zero horizontal dispersion
    where damping wigglers are located are the L0 straight ($s=0~\pm$ 10~m where
    there are 6 wigglers) and
    L1 (126.7~m to 132.7~m, 3~wigglers) and L5 ($s=635.8$~m to $s=641.8$~m, 3~
    wigglers). L3 ($384.2~\pm 9$~m) is the experimental straight that
    includes the chicane for electron cloud studies and the in-situ SEY
    station. Retractable mirrors are used to reflect synchrotron
    radiation generated in the bends on either side of the L3 straight
    to the cave where optical instruments are located.  }
\end{figure}

\begin{figure}[tb] 
   \centering
   \includegraphics[width=4in]{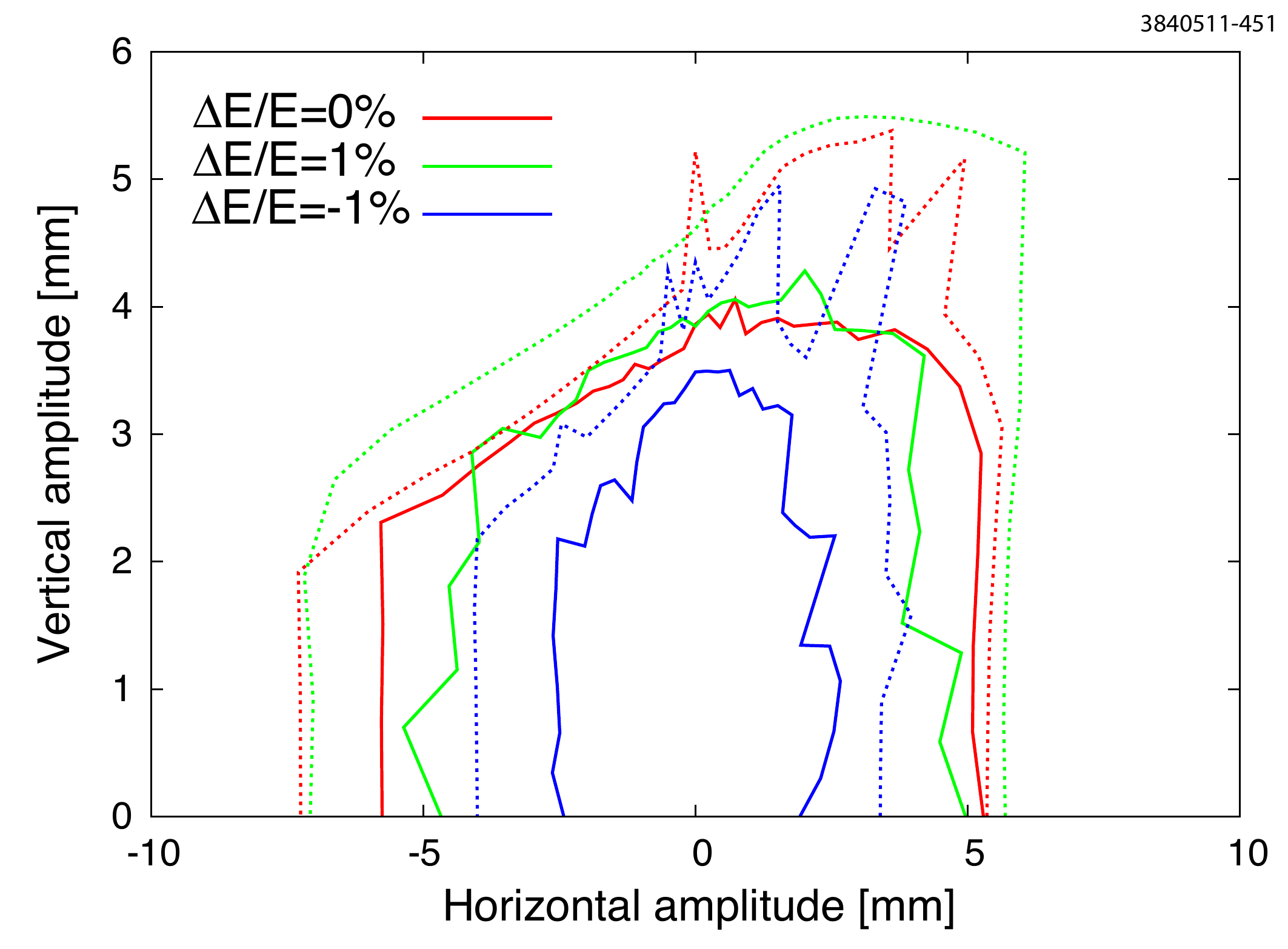}
  \caption[The dynamic aperture for the low emittance lattice at
  2.085GeV.]
{\label{fig:intro:da} Simulation of the dynamic aperture for
  the low emittance lattice at 2.085GeV. The solid curves are the maximum
    initial amplitude that survives 1000 turns at betatron tunes $Q_h~=~14.595$,
    $Q_v~=~9.63$, and synchrotron tune $Q_z~=~0.0645$. The red, green and blue correspond
    to initial energy offsets of $0,\pm1\%$. The dashed lines are the
    maximum initial amplitude that survives for 20 turns and so
    approximate the physical aperture. Particles are lost at the real
    physical boundaries of the vacuum chamber.
}
\end{figure}

CESR achieves minimum emittance at a 2~GeV beam energy by employing
superconducting damping wigglers, having a main period of 40 cm and operating at 
1.9~T field strength.\cite{PAC03:TOAB007, PAC03:WPAE009, PAC07:MOZBKI01, PAC07:THPMS011} The wigglers
reduce the radiation damping time by an order of magnitude and the
horizontal emittance by a factor of 5 with respect to the wiggler off condition. The lattice functions are
arranged so that there is zero horizontal dispersion in the wiggler
straights. The relatively strong focusing and high horizontal tune minimizes the dispersion
in the bend magnets and the resulting horizontal emittance. The
lattice functions for the minimum emittance 2.085~GeV optics are shown
in Figure~\ref{fig:intro:magnetic_optics}. The independently-powered sextupoles are deployed to
correct chromaticity in two families. Dynamic aperture is shown in
Figure~\ref{fig:intro:da}.

The damping wigglers are essential to achieving the low emittance in
these optics. However, in the limit of a
wiggler dominated ring, the minimum emittance is determined by the
horizontal emittance generated within the wigglers themselves. With 12
wigglers that minimum is indeed obtained with a 1.9 T wiggler
field. Higher wiggler fields will further reduce the damping time, but
will increase the emittance due to the wiggler dispersion. The only
way to further reduce the emittance with wigglers would be to increase
the total length of wigglers, rather than increasing the wiggler
field.

\subsection{Energy Reach}
We have designed and commissioned optics at 1.8, 2.085, 2.3 and
2.8~GeV, with all wigglers at 1.9~T field. At beam energy greater than
3~GeV, wiggler operation is limited to the those in the L0
straight. The vacuum chambers adjacent to the arc wigglers cannot
sustain the synchrotron radiation power generated by these wigglers at high energies.
Table \ref{tab:intro:lattices} lists some of the {\cesrta} lattice
configurations that have been used for various measurements.  The column,
labelled Wiggler rad, refers to the fraction of the synchrotron radiation for the
storage ring, which is produced by the wigglers.  The rightmost column gives
the fractional energy spread $\sigma_E$/E with the superconducting wigglers
on and in optics roughly equivalent with the wigglers off.  N.b. since the wigglers
naturally provide vertical focusing, turning off wigglers requires significant
changes in the quadrupole strengths near the wigglers.  Although the wiggler off
optics is not identical to the wiggler on optics, the energy spread variation caused
by the wigglers being off is very nearly as specified in Table \ref{tab:intro:lattices}.

\begin{table}[htb]
\begin{center}
\caption[Initial \cesrta Lattice Configurations]{\label{tab:intro:lattices}Typical {\cesrta} Lattice Configurations} \vspace*{1ex}
\begin{tabular}{|l|c|c|c|c|c|c|c|}
\hline
Lattice Name &  Energy & Emittance & Wigglers & $\beta_v$
at & Wiggler & $\sigma_E$/E  \\
 &  [GeV] &[nm-rad] & @ 1.9~T & xBSM & rad [\%] & (with/without) \\
 & & & & $ e^+$~source & & Wigglers [\%]  \\
\hline
 CTA\_1800MEV\_ - &1.8&2.0&12      &  40 & 90 & (0.0767~/  \\
 ~XR40M\_20110520 &~&~&~&~&~& 0.0210) \\
 CTA\_2085MEV\_ - &2.085&2.6&12      &  20& 87 & (0.0813~/  \\
 ~XR20M\_20091205 &~&~&~&~&~& 0.0244) \\
 CTA\_2085MEV\_ - &2.085&2.6&12      &  40 & 87 & (0.0813~/  \\
 ~XR40M\_20091205 &~&~&~&~&~& 0.0244) \\
 CTA\_2085MEV\_ - &2.085&2.6&12      &  5.8 & 87 & (0.0813~/  \\
 ~20090516 &~&~&~&~&~& 0.0244) \\
 CTA\_2300MEV\_ - &2.3&3.2&12      & 40& 84 & (0.0843~/  \\
 ~XR40M\_20110531 &~&~&~&~&~& 0.0269) \\
 CTA\_3000MEV\_ - & 3.0 & 10.0 & 6 & 11.4 & 58 & (0.0841~/  \\
 ~Q0H\_20090822 &~&~&~&~&~& 0.0350) \\
 CTA\_4000MEV\_ - & 4.0 & 23 & 6 & 10.7&47 & (0.0888~/  \\
 ~23NM\_20090816 &~&~&~&~&~& 0.0467) \\
 CTA\_5000MEV\_ - & 5.0 & 40 & 6 & 7.2&37 & (0.0903~/  \\
 ~40NM\_20090513 &~&~&~&~&~& 0.0584) \\
 CTA\_5000MEV\_ - & 5.0 & 74 & 0& 11.3&0 & (~~--~~/  \\
 ~20090311&~&~&~&~&~& 0.0584) \\
\hline
\end{tabular}
\end{center}
\end{table}

The use of superconducting wigglers also has an impact for the maximum beam current at high energy.  The maximum wiggler power incident on the synchrotron radiation absorbers is limited to 40~kW for the six wigglers used at high energies.  This limit was determined by synchrotron radiation calculations of the power deposited onto the absorbers with an opening angle of {1/$\gamma$} at a fixed beam current.  This information was processed using ANSYS to calculate a temperature and stress distribution in the absorbers. We have chosen to limit the beam current above 4~GeV beam energies to maintain a safety factor of at least three between the calculated maximum stress and the material's yield stress.  The maximum operating currents for CESR at various operating energies is given in Table~\ref{tab:intro:currents}.  The very low current limits for electrons above 4 GeV with wigglers powered is due the lack of high power absorbers downstream of the L0 straight for the electron direction.  More details on the superconducting wigglers may be found in Part 2, which describes the vacuum system.

\begin{table}[htb]
\begin{center}
\caption[\cesrta Operating Current Limits]{\label{tab:intro:currents}{\cesrta} Operating Current Limits.  These are the administrative limits placed on beam currents at different operating energies.} \vspace*{1ex}
\begin{tabular}{|c|c|c|c|c|}
\hline
Beam Energy &  Maximum Positron & Maximum Electron & Limited by \\
$$ [GeV] & Total Current & Total Current & Wiggler Power \\
  &[ mA] & [mA] &  \\
\hline
 1.8 & 150 & 100 & no \\
 2.085 & 150 & 100 & no \\
 2.3 & 150 & 100 & no \\
 2.5 & 150 & 100 & no \\
 3.0 & 150 & 10 & yes \\
 4.0 & 100 & 3 & yes \\
 5.0 & 40 & 1 & yes \\
 5.0 & 240 & 240 & no \\
 5.3 & 240 & 240 & no \\
\hline
\end{tabular}
\end{center}
\end{table}

\subsection{Grouped Control System Elements}
\begin{figure}[tb] 
   \centering
   \includegraphics[width=5in]{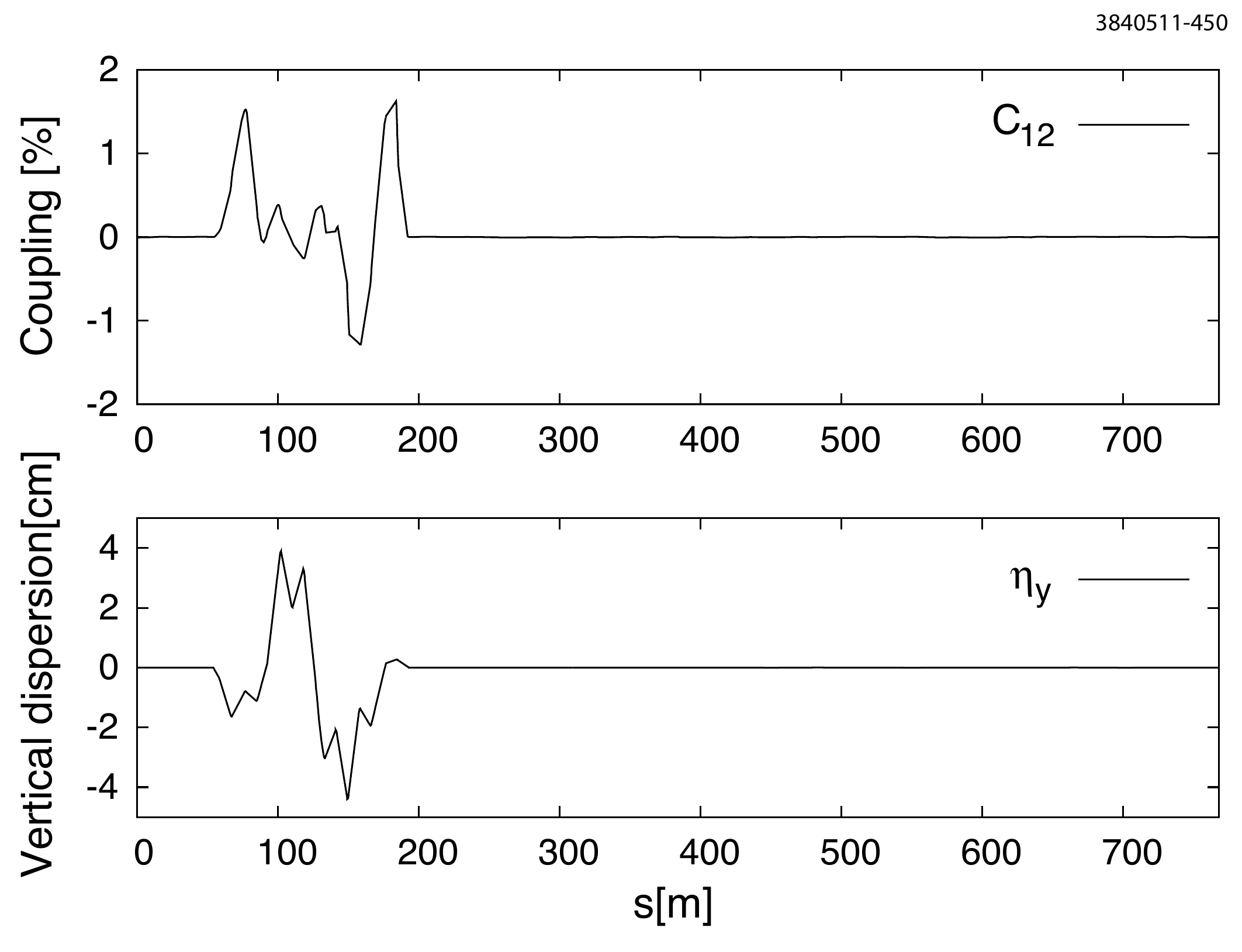}
  \caption[Closed coupling and vertical dispersion bump is generated with seven skew quads]{\label{fig:intro:bumps} The closed coupling and vertical
    dispersion bump is generated with seven skew quads. The
    theoretical change to the vertical emittance (assuming that the
    unperturbed machine has zero coupling and zero vertical
    dispersion), is 25~pm-rad. The dispersion peak appears in the L1
    region of the west arc wiggler straight. Bumps like this one are used to
    characterize the vertical beam size monitor and to vary emittance
    for investigation of intra-beam scattering, ion, and electron cloud effect.  }
\end{figure}

The flexibility of the CESR control system and the independent powering of
all CESR magnets allow the possibility of creating grouped control elements 
(also called control knobs) for closed orbit, $\beta$, dispersion,
and coupling bumps.  We use orbit bumps to align the beam to position the source points for
x-ray beam lines as required for the x-ray beam size monitor. In general, the beam size
$\sigma$ depends on emittance $\epsilon$, amplitude function $\beta$, dispersion $\eta$,
and energy spread $\delta$, according to
$\sigma = \sqrt{\epsilon\beta+(\eta^2\delta^2)}$. It is often
convenient to be able to independently
vary $\beta$, $\epsilon$ and dispersion to explore beam size monitor systematics and properties of the lattice.
$\beta$-bumps are used to
manipulate the beta function at the x-ray and visible light beam size monitor
source points.  A closed dispersion bump at the source of the
horizontal beam size monitor allows a measurement of the contribution
of the energy spread to the beam size. Closed coupling/dispersion bumps are used to vary
vertical dispersion (and therefore vertical emittance) in a controlled
way. We have developed code, which automatically computes coefficients and
loads
into the control system data base complete sets of orbit and
coupling/dispersion bumps for each new lattice configuration.
An example of the effect of such a coupling/dispersion bump is shown
in Figure~\ref{fig:intro:bumps}.


\section[General Accelerator Modifications and Upgrades]{General Accelerator Modifications and Upgrades}
\label{sec:cesr_conversion.gen_mod}

\subsection{Overview}
\label{ssec:cesr_conversion.vac_system.overview}
  The CESR storage ring, shown in figure~\ref{fig:cesr_conversion:vac_fig1}, is capable of storing two counter-rotating beams with total currents up to 500~mA (or a single beam up to 250 mA) at a beam energy of 5.3~GeV.  As shown in Table \ref{tab:intro:parameters} the storage ring has a total length of 768.44~m, consisting of primarily bending magnets and quadrupoles in the arcs, two long straight sections, namely L0 (18.01~m in length) and L3 (17.94~m in length), and four medium length straights (namely, $L1, L5,$ both 8.39~m in length and $L2, L4,$ both 7.29~m in length).

\subsection{Wiggler Straight Section Reconfiguration}
\label{ssec:cesr_conversion.gen_mod.wiggler}

\begin{figure}[htbp] 
   \centering
   \includegraphics[width=0.95\columnwidth]{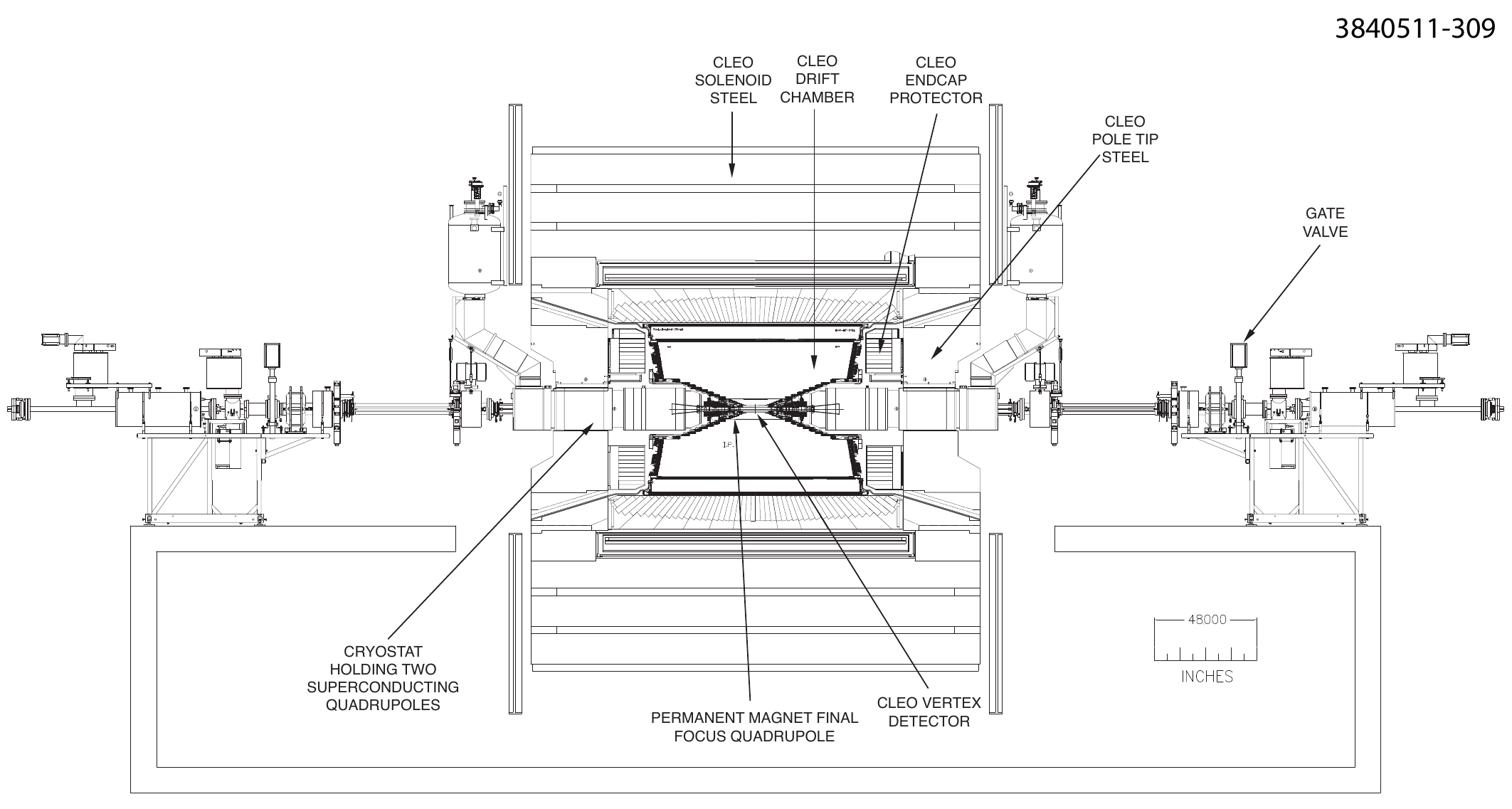}
   \caption[Elevation view of the CESR-c/CLEO-c interaction region]{\label{fig:L0OriginalLayout}
   An elevation view of the CESR-c/CLEO-c interaction region before reconfiguration as the wiggler straight section for \cesrta.  The central section of the CLEO-c detector, the final focusing superconducting quadrupoles and the connecting vacuum chambers were removed during the {\cesrta} installation. }
\end{figure}

CESR required extensive modifications for the straight section that
contained the CLEO-c high energy physics (HEP) experimental
detector\cite{PAC09:FR1RAI02}.  During HEP operation this straight
section was a micro-beta insert utilizing four superconducting
quadrupoles and two permanent magnet final focus quadrupoles, all of
which were oriented with approximately 4$^\circ$~tilts to compensate
the CLEO solenoidal magnetic field.  There was an additional pair of
skew quadrupoles within the interaction region straight section to
complete the solenoid compensation.  During HEP operations the
electron and positron bunches crossed at the interaction point with
approximately a $\pm$2~mrad crossing angles, created by four
horizontal separators placed symmetrically in the arcs of CESR.  The
layout of the HEP interaction region straight section is displayed
in figure~\ref{fig:L0OriginalLayout}.

\begin{figure}[htb] 
    \centering
    \includegraphics[width=0.75\textwidth]{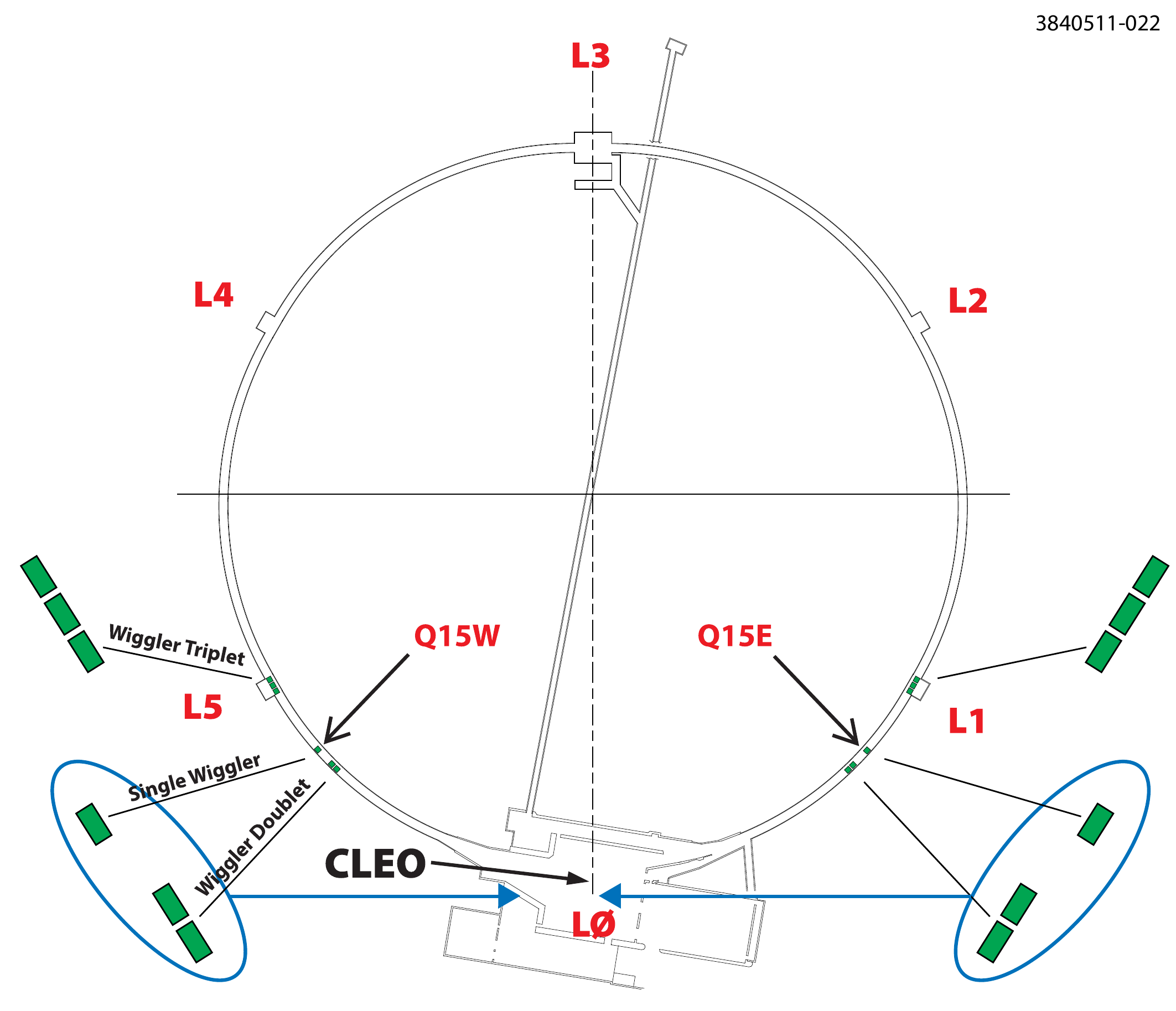}
    \caption[{\cesrta} Vacuum System]{The reconfiguration of  CESR accelerator components provided space in two long regions in L0 and L3, and two flexible short regions at Q15W and Q15E. Hardware for electron cloud studies was installed in these regions. \label{fig:cesr_conversion:vac_fig1}}
\end{figure}

As a part of the CESR-c/CLEO-c HEP program twelve superconducting
wigglers (SCWs) were installed in the southern one third of
CESR\cite{PAC07:MOZBKI01}.  For 2.1~GeV operation these wigglers
provided 90\% of the radiation damping in CESR and in their original
arc locations could be used for emittance control of the colliding
beams.  During CESR-c/CLEO-c HEP operations 6 of the 12~SCWs were
installed as two triplet SCWs, located at two straight sections,
namely L1 and L5, and the remaining 6~SCWs were in shorter straight
sections between L0 and L1, and between L0 and L5.  For the lowest
emittance operation for {\cesrta} all twelve wigglers must be
located in regions with zero dispersion. The {\cesrta} lattice
provides for zero dispersion regions in the L0, L1 and L5 straight
sections, which are shown in
figure~\ref{fig:cesr_conversion:vac_fig1}.  Therefore six of the
CESR-c wigglers were relocated to the CESR L0~straight, in the place
of the CLEO drift chamber and endcaps.  The other six wigglers,
located in the L1 and L5 straight sections, remained in their
original places.

\begin{figure}[hbt]
    \centering
    \includegraphics[width=\textwidth]{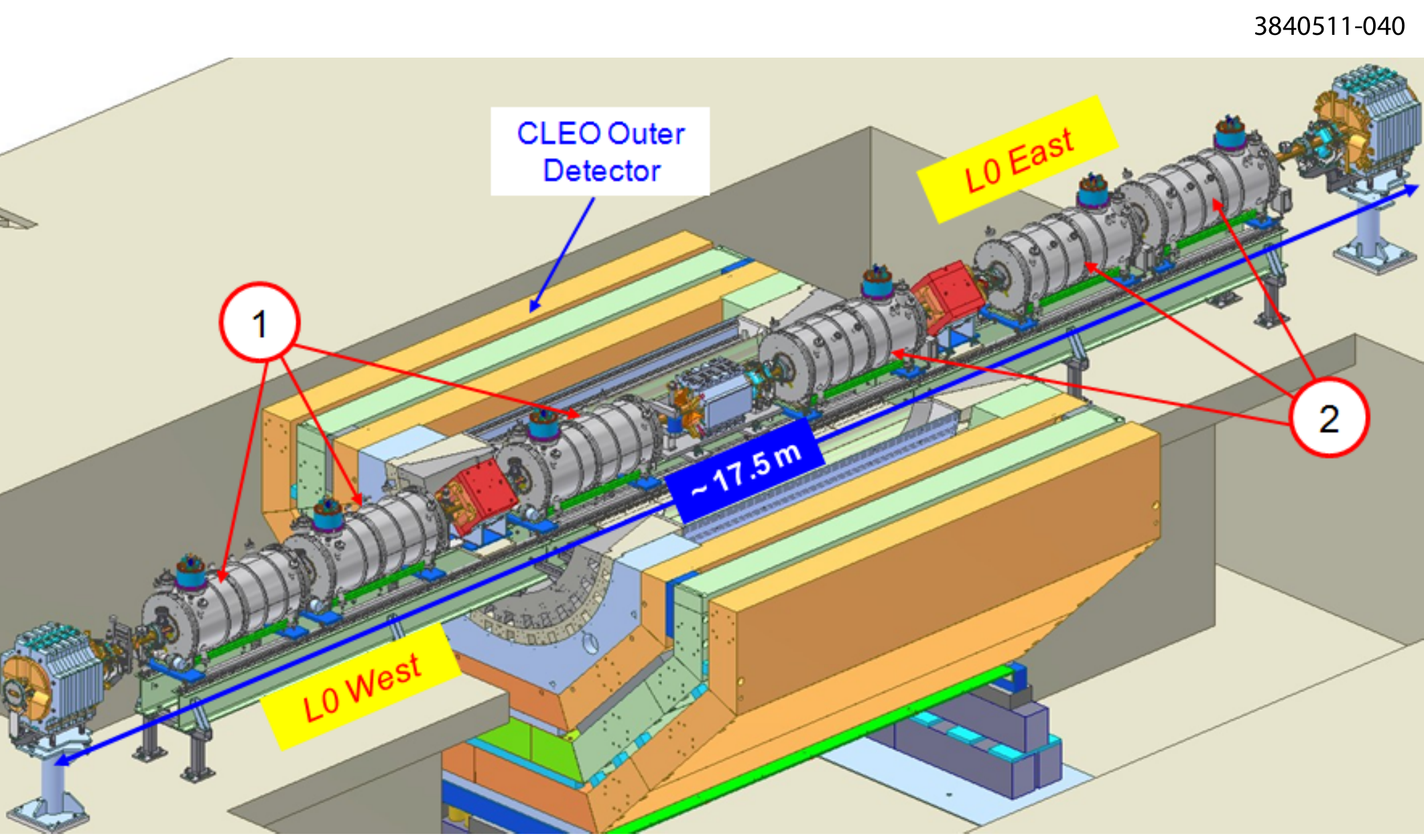}
    \caption[L0 EC Experimental Region Vacuum Layout]{L0 center {\cesrta} EC experimental region, consists of (1) three retarding field analyzer-equipped SCWs and (2) three CESR-c SCWs.  Many other EC diagnostics, such as retarding field analyzer (RFA) in the drifts, BPMs and TE-Wave buttons, are also implemented.\label{fig:cesr_conversion:vac_l0_center}}
\end{figure}

\begin{figure}[htbp] 
   \centering
   \includegraphics[width=0.8\columnwidth]{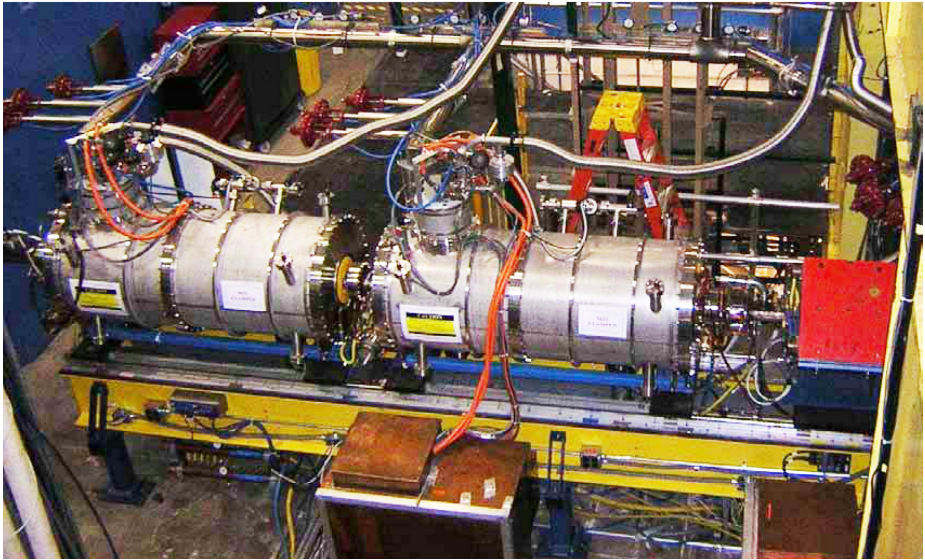}
   \caption[View of the wiggler straight section from the Northeast.]{\label{fig:L0Beamline1a}
   View of the wiggler straight section from the North on the East side of the wiggler straight section during the operation of {\cesrta}.  Two of the wiggler cryostats are visible as is a part of Q01E (orange magnet to the right of the cryostats.)  }
\end{figure}

During the July~2008 shutdown, the central portion of the CLEO detector was decommissioned by removing the superconducting, normal conducting and permanent quadrupoles, the steering magnets for CESR, the CLEO endcap detectors, the vertex detector, the drift chamber and all of their associated cabling along with approximately 17~meters of vacuum chambers.  All of these were located between the soft bend dipole magnets.  A pair of bridging I-beams was installed through CLEO iron to support the quadrupole and steering magnets and the SCWs, which were positioned in the central portion of the L0 straight section. The six SCWs originally in the short straight sections between L0-L1 and L0-L5 were the wigglers relocated to the L0~long straight section as shown in figure~\ref{fig:cesr_conversion:vac_l0_center}.  With the relocation all 12~SCWs are positioned in the long straight sections, in which the optics can be configured for zero dispersion, to produce the smallest possible beam emitances.  Figure~\ref{fig:cesr_conversion:vac_l0_center} shows the beam pipe, the conventional magnets and the six SCWs placed within the vacated CLEO detector's solenoid iron yoke.  Figure~\ref{fig:L0Beamline1a} displays a view of the wiggler straight section.

\subsection{L3 Straight Section Reconfiguration}
\label{ssec:cesr_conversion.gen_mod.l3_straight}

\begin{figure}[htbp] 
   \centering
   \includegraphics[width=0.95\columnwidth]{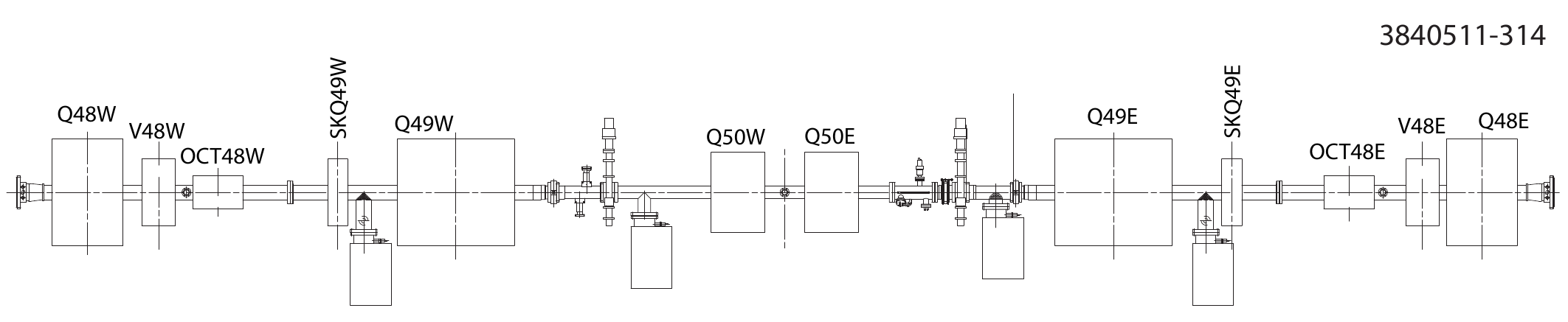}
   \caption[L3 optics between the vertical separators before reconfiguration for {\cesrta}.]{\label{fig:L3OriginalLayout1}
   Schematic layout of the L3 optics between the vertical separators before reconfiguration for {\cesrta} operations.  }
\end{figure}

\begin{figure}[htbp] 
   \centering
   \includegraphics[width=0.95\columnwidth]{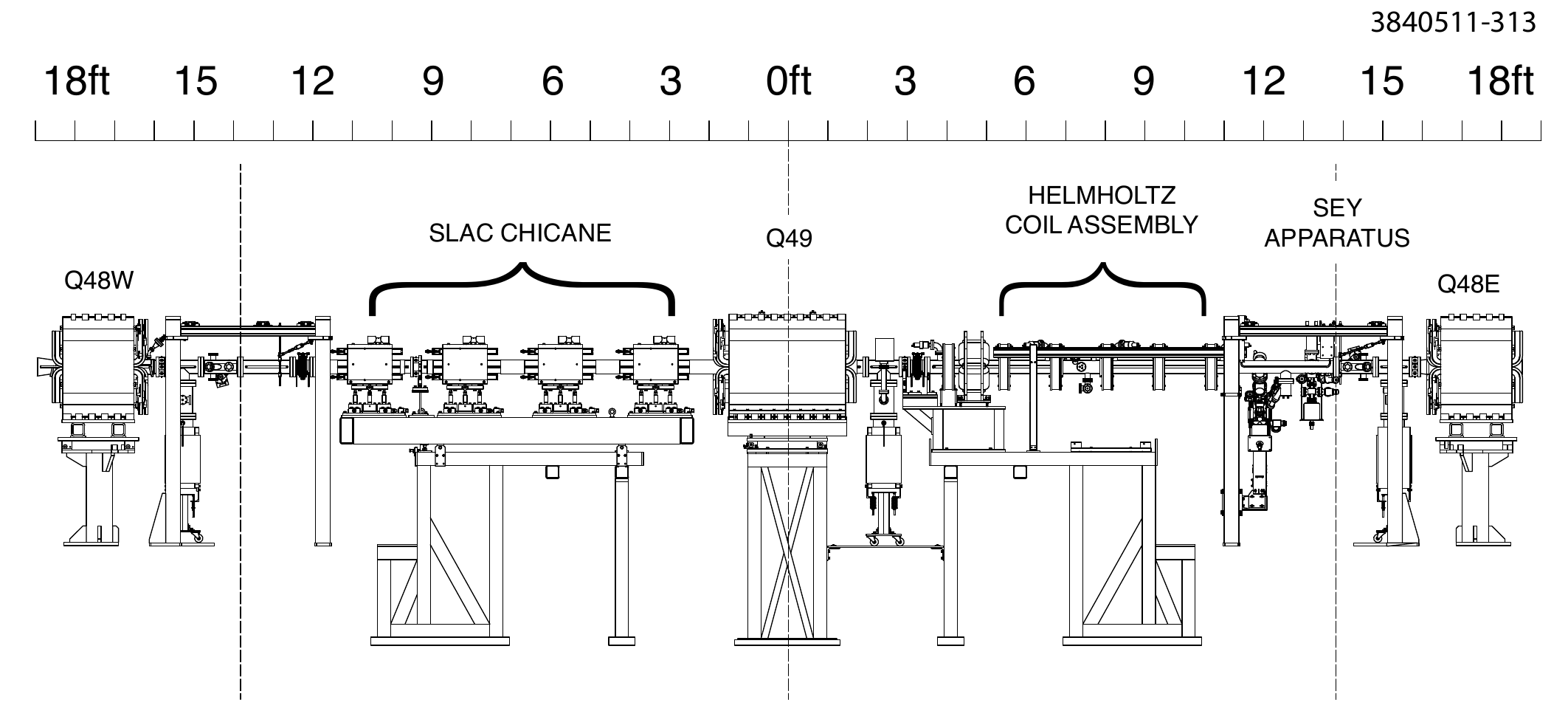}
   \caption[Schematic layout of the L3 experimental region.]{\label{fig:L3FinalLayout}
   Schematic layout of the L3 experimental region.  The experimental area includes the following vacuum chamber test regions: a chicane section, a region that has Helmholtz coils around the beam pipe to allow for chamber processing via bakeout etc., and an SEY apparatus with a lock-load mechanism to permit easy access for changing vacuum chamber wall surfaces.  }
\end{figure}

\begin{figure}[htbp] 
   \centering
   \includegraphics[width=0.7\columnwidth]{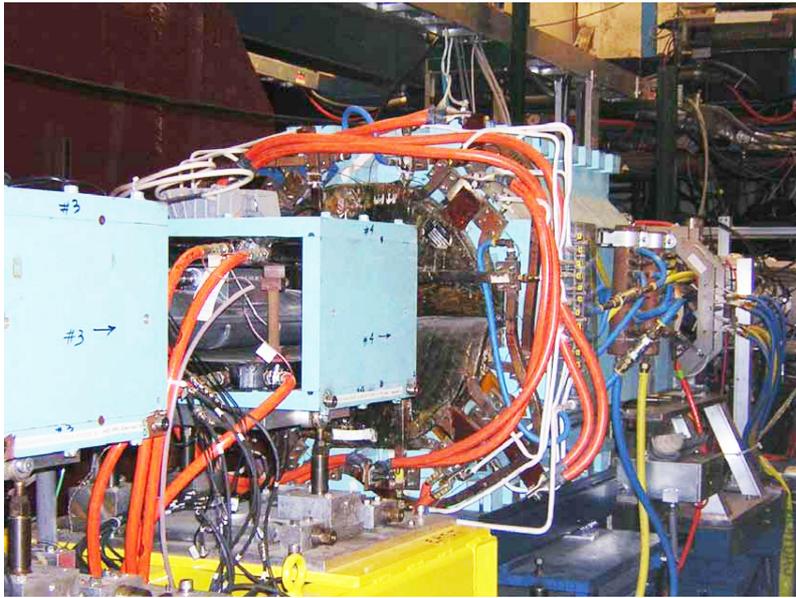}
   \caption[Picture of the L3 experimental region looking to the East .]{\label{fig:L3Beamline1}
   View of the L3 experimental region looking to the East and showing two of the chicane magnets in the foreground followed by Q49. }
\end{figure}

\begin{figure}[htbp] 
   \centering
   \includegraphics[width=0.7\columnwidth]{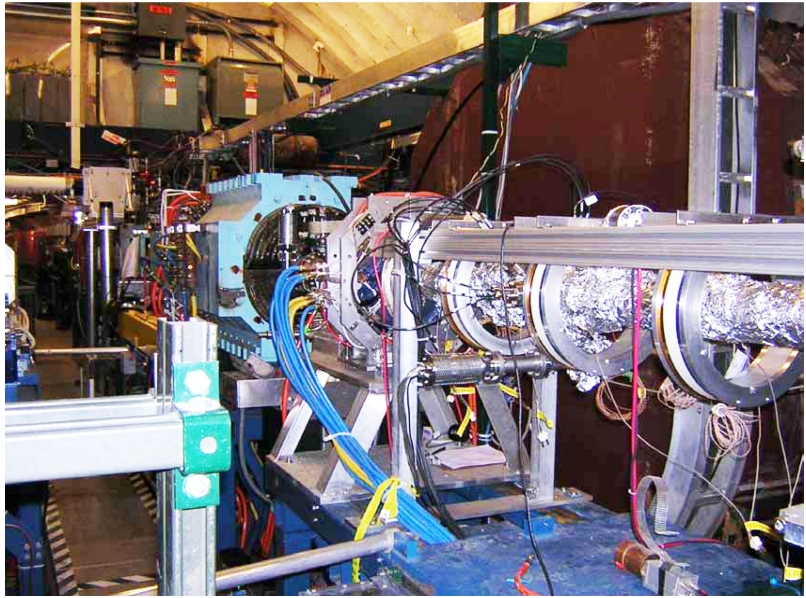}
   \caption[L3 experimental region viewing Q49 from the East.]{\label{fig:L3Beamline2}
   L3 experimental region viewing Q49 from the East and having the Helmholtz coils around the beam pipe for studies of Electron Cloud suppression.   The picture also shows the Helmholtz coils surrounding the vacuum chamber as it is being prepared for a bakeout. }
\end{figure}

During CESR-c/CLEO-c HEP operations the L3 straight section (diametrically opposite to the CLEO-c detector's straight section) was configured to have a pair of electrostatic vertical separators, necessary to separate electron and positron bunches at the second horizontal angle crossing point for CESR, and an additional six quadrupoles forming a mini-beta insert.  A schematic of the original optics layout for the central region of the L3 straight section is found in figure~\ref{fig:L3OriginalLayout1}.  To accommodate the planned experimental regions in this straight section, a major change to the accelerator optics was undertaken.  After the removal of the pair of electrostatic vertical separators (outboard of the Q48W and Q48E quadrupoles), a long experimental straight section was established in the north region of CESR (L3).  This 12-meter section, as shown in figure~\ref{fig:L3FinalLayout}, is currently hosting many SLAC~EC~beam pipes for study and diagnostics, including a set of 4-dipole chicane magnets with beam pipes equipped with EC detectors, and an aluminum beam pipe with grooved interior. A pair of retractable synchrotron-light mirrors (highly polished beryllium) are set up for diagnostics and are used for beam profile measurements.  Two in situ secondary electron emission yield (SEY) measurement systems were installed in the sector \cite{ARXIV:1412.3477}.  The SEY systems are equipped with load-locks, so SEY may be measured as a function of the beam dose for commonly used vacuum materials.  Pictures of the experimental region are found in figure~\ref{fig:L3Beamline1} and figure~\ref{fig:L3Beamline2}.

\subsection{Solenoid Windings}
\label{ssec:cesr_conversion.gen_mod.solenoids}

\begin{figure}[htbp] 
   \centering
   \includegraphics[width=0.6\columnwidth]{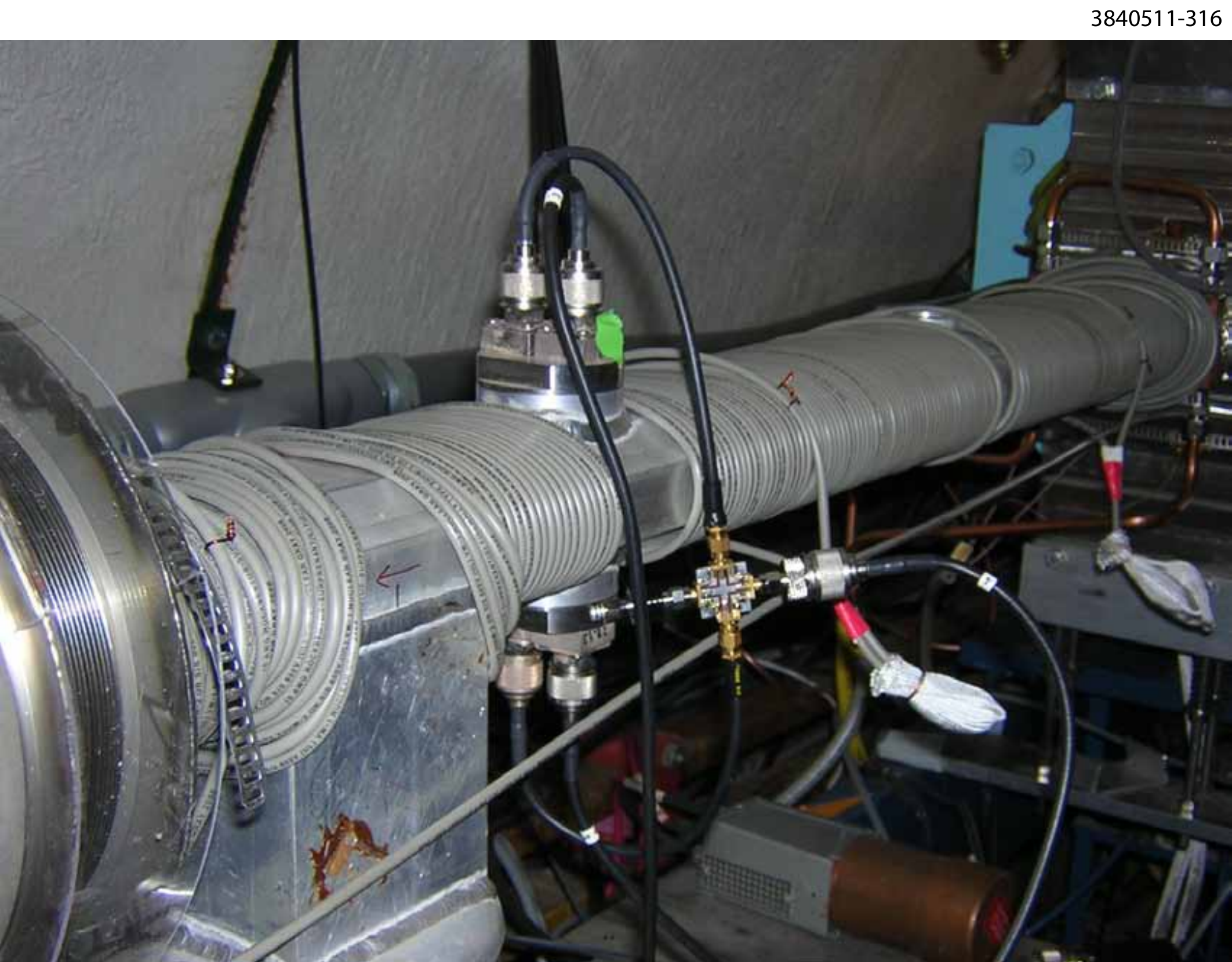}
   \caption[View of solenoid windings on one typical arc vacuum chamber.]{\label{fig:SolenoidWinding1}
   View of solenoid windings on one typical arc vacuum chamber. }
\end{figure}

\begin{figure}[htbp] 
   \centering
   \includegraphics[width=0.6\columnwidth]{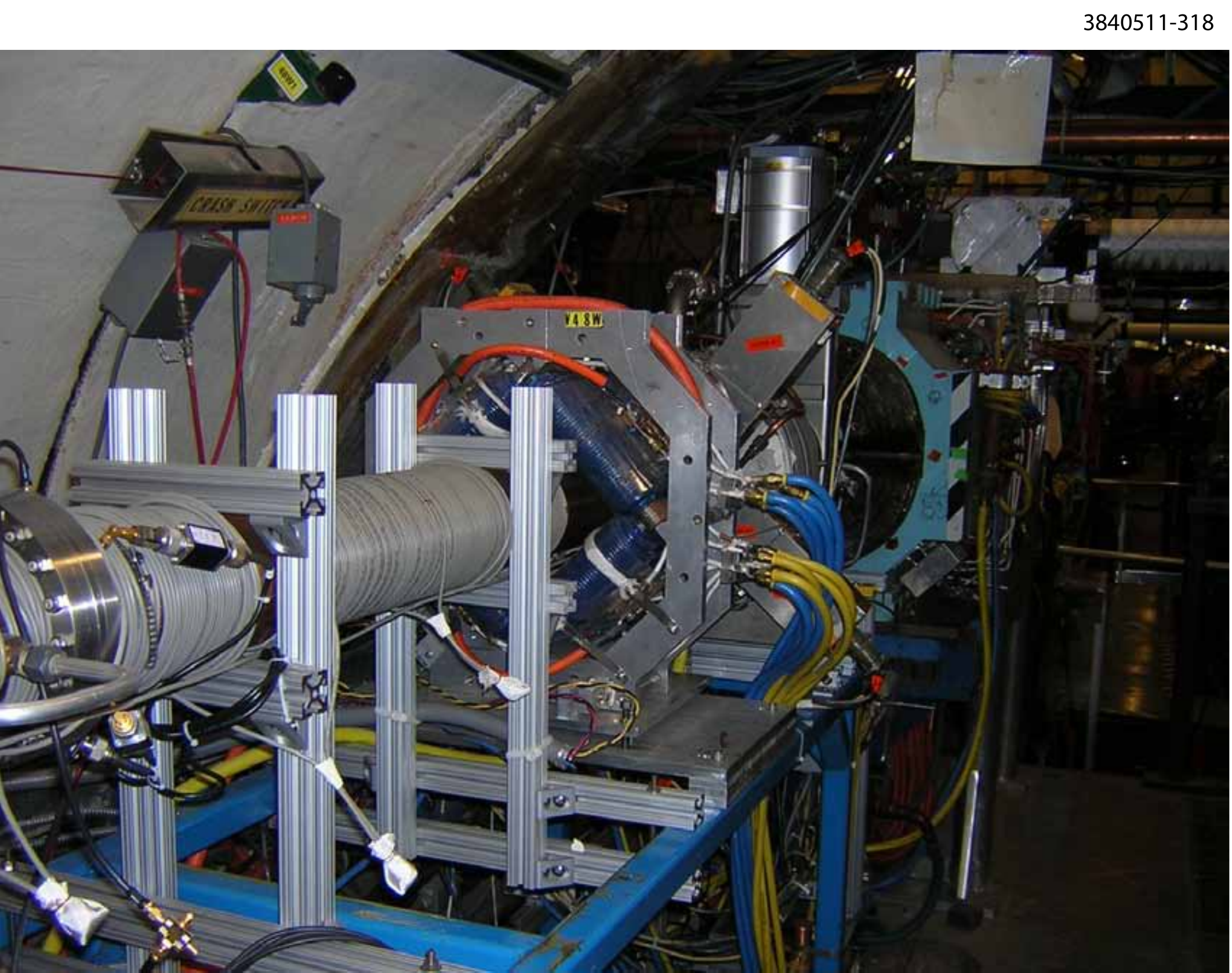}
   \caption[View of solenoidal field windings on the L3 straight section vacuum chambers.]{\label{fig:SolenoidWinding3}
   View of solenoidal field windings on the L3 straight section vacuum chambers.}
\end{figure}

Solenoid windings on drift sections of storage rings have been
successfully employed to reduce the effect of electron 
clouds.\cite{PRSTAB7:024402,ICFABDNL48:112to118}.  Although
the total drift section length of CESR is only approximately 15\% of
the circumference and was not expected to play a major role in the
electron cloud dynamical effects for the entire ring, elliptical
solenoid windings have been added to cover approximately 80\% of
this drift length.  The windings were wrapped directly on the CESR
vacuum chamber after a thin Kapton insulating layer was added around
the radial outside of the chamber to effectively enlarge the 3~mm radius 
corners above and below the water cooling channel to prevent
the radii of curvature for the solenoid cable becoming too small at these
two corners.  In 
a few sections of the storage ring the beam pipe is
circular so in these places the windings are cylindrical solenoids.
In the experimental region in the L3~straight section, several
Helmholtz coils were employed to allow better access to the vacuum
chamber; the spacing between the coils was set to have the
longitudinal field for these coils approximately the same as in the
standard solenoid windings.  The cable, employed for the windings
wrapped directly on the beam pipe, is radiation hard number 10~(AWC)
gauge insulated wire, which is wound in a single close-packed single
layer.  A few examples of the Helmholtz coils and solenoid windings
maybe seen in figure~\ref{fig:L3Beamline2},
figure~\ref{fig:SolenoidWinding1} and figure~\ref{fig:SolenoidWinding3}.
Windings in adjacent drift sections are clustered together and
connected in series to one switching DC~power supply, where an
effort has been made to reduce the net longitudinal magnetic field
by arranging the polarity of the windings to roughly balance the
number of turns which have the current flowing clockwise around the
vacuum chamber with those where it is flowing counter-clockwise.
This reduces the local horizontal-vertical coupling of the beam's
motion caused by each supply powering an individual cluster of
solenoid windings.

The power supplies are generally unipolar 25~A switching DC~power supplies operating off of the common 65~VDC ring magnet power supply bus.  The power supplies function with the same control system hardware and software as the CESR steering power supplies.  The winding resistance for the adjacent drift sections, as clustered together and connected in series, is less than 2.5~$\Omega$, being low enough to allow the full 25~A of current.  In two instances where the solenoid windings surround the shielded pickups, 100~A~bipolar supplies were installed (and limited to $\pm$~25~A operation) to permit sweeping the solenoid field over both polarities for the shielded pickup measurements.

During installation the polarities of the windings for all of the separate drift sections was verified to have the required alternation needed to reduce the horizontal-vertical coupling effects.  After installation the magnitude of the longitudinal magnetic field was measured to be approximately 40~G for a 25~A excitation in the standard CESR beam pipe windings.  A single positron bunch in 2.1~GeV conditions was used to check the coupling error caused by each of the 16~clusters of windings.  After adjusting the minimum tune split on the coupling resonance to be less than 0.0002, the change in the global coupling of the bunch was measured when all of the solenoid power supplies were excited to their full currents. With all solenoid power supplies at full current the accelerator minimum tune split on the coupling resonance was 0.0027.  Since the solenoids were installed primarily to study their effect on the cloud's density, they are usually powered only during mitigation or shielded pickup measurements.

\subsection{Magnet System Controls Upgrades}
\label{ssec:cesr_conversion.gen_mod.controls_software}

The conversion of CESR to a test accelerator to study low emittance beams and the effects of electron clouds has required significant additions or changes to the accelerator controls.  The first set of changes is obvious from the preceding sections: the reconfiguration of the magnet controls for the conventional accelerator magnet system.  The change to CESR's optics and steering controls required the removal, installation, reconfiguration or re-cabling of 48~separate magnets in the ring for the wiggler and L0 straight sections.  The elements, which were installed for {\cesrta} operations, are summarized in Table~\ref{tab:MagnetPS}.

\begin{table}[htpb]
   \centering
   \caption{\label{tab:MagnetPS} Listing of standard CESR accelerator elements, which were installed, re-cabled or reconfigured for {\cesrta} operation.}
   \vspace*{1ex}
   \begin{tabular}{|c|c|c|c|c|}
   \hline\hline
   Name(s) of & Type of & Maximum & Maximum & Number \\
   {\cesrta} Element & Power Supply & Current & Voltage & of PSs\\
   \hline
   Q00W & Linear Pass-bank & 250 A & 28 V & 1 \\
    & Regulated & & & \\
   \hline
   Q01W, Q01E & Precision Chopper & 80 A & 55 V & 2 \\
    & Switching Regulator & & & \\
   \hline
   Q02W (16T), & Precision Regulated & 1000 A & 20 V & 2 \\
   Q02E (16T) & EMI PS & & & \\
   \hline
   Q48W (6T PS) & In series with & 700 A & 300 V & 1  \\
   Q48E (6T PS) & dipole magnets & & &  \\
   & - Transrex PS's & & &  \\
   \hline
   Q48W (22T), & Linear Pass-bank & 250 A & 28 V & 2 \\
   Q48E (22T) & Regulated & & & \\
   \hline
   Q49W (22T) & Precision Regulated & 1000 A & 20 V & 1 \\
    & EMI PS & & & \\   \hline
   H01W & Bipolar Chopper & $\pm$12.5 A & $\pm$55 V & 1 \\
    & Switching Regulator & & & \\
   \hline
   V01E & Bipolar Chopper & $\pm$12.5 A & $\pm$55 V & 1 \\
    & Switching Regulator & & & \\
   \hline
   V02W, V02E & Bipolar Chopper & $\pm$12.5 A & $\pm$55 V & 2 \\
    & Switching Regulator & & & \\
   \hline
   SQ01E & Bipolar Chopper & $\pm$12.5 A & $\pm$55 V & 1 \\
    & Switching Regulator & & & \\
   \hline
   SQ02W,  & Bipolar Chopper & $\pm$12.5 A & $\pm$55 V & 2 \\
    SQ02E & Switching Regulator & & & \\
   \hline
   H49E & Bipolar Chopper & $\pm$12.5 A & $\pm$55 V & 1 \\
    & Switching Regulator & & & \\
   \hline
   V49E & Bipolar Chopper & $\pm$12.5 A & $\pm$55 V & 1 \\
    & Switching Regulator & & & \\
   \hline
   V48W, V48E & Bipolar Chopper & $\pm$12.5 A & $\pm$55 V & 2 \\
    & Switching Regulator & & & \\
   \hline
   SQ48W, SQ48E & Bipolar Chopper & $\pm$12.5 A & $\pm$55 V & 2 \\
    & Switching Regulator & & & \\
   \hline \hline
   \end{tabular}
\end{table}

\begin{figure}[htbp] 
   \centering
   \includegraphics[width=0.6\columnwidth]{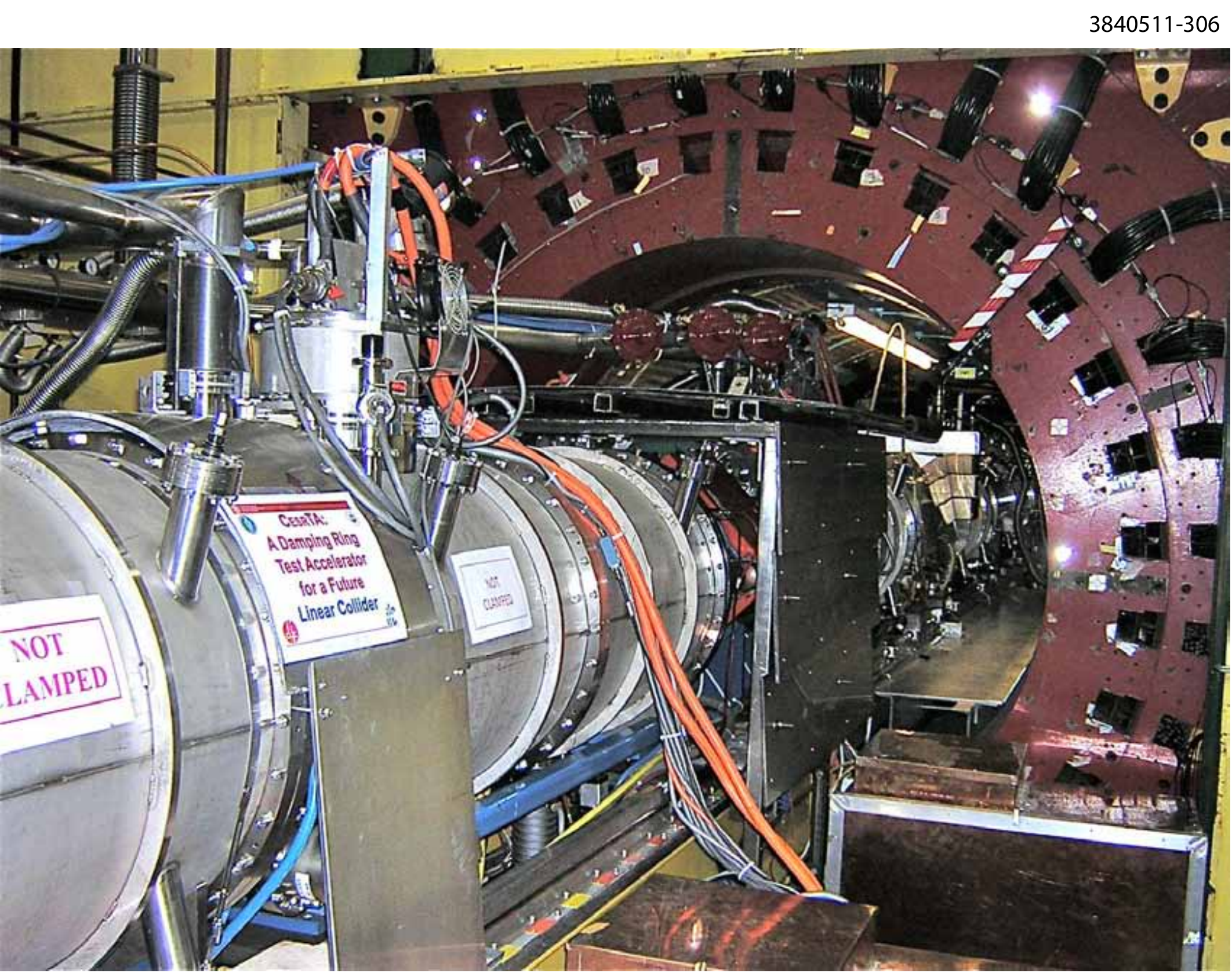}
   \caption[Wigglers in the wiggler straight section.]{\label{fig:CtaWigglerControls2}
   Wigglers in the wiggler straight section with their controls.  The picture shows a view looking into the former CLEO detector with two of the three wigglers displayed to the left and the cabling of rightmost of these two in the center of the picture.}
\end{figure}

\begin{figure}[htbp] 
   \centering
   \includegraphics[width=0.6\columnwidth]{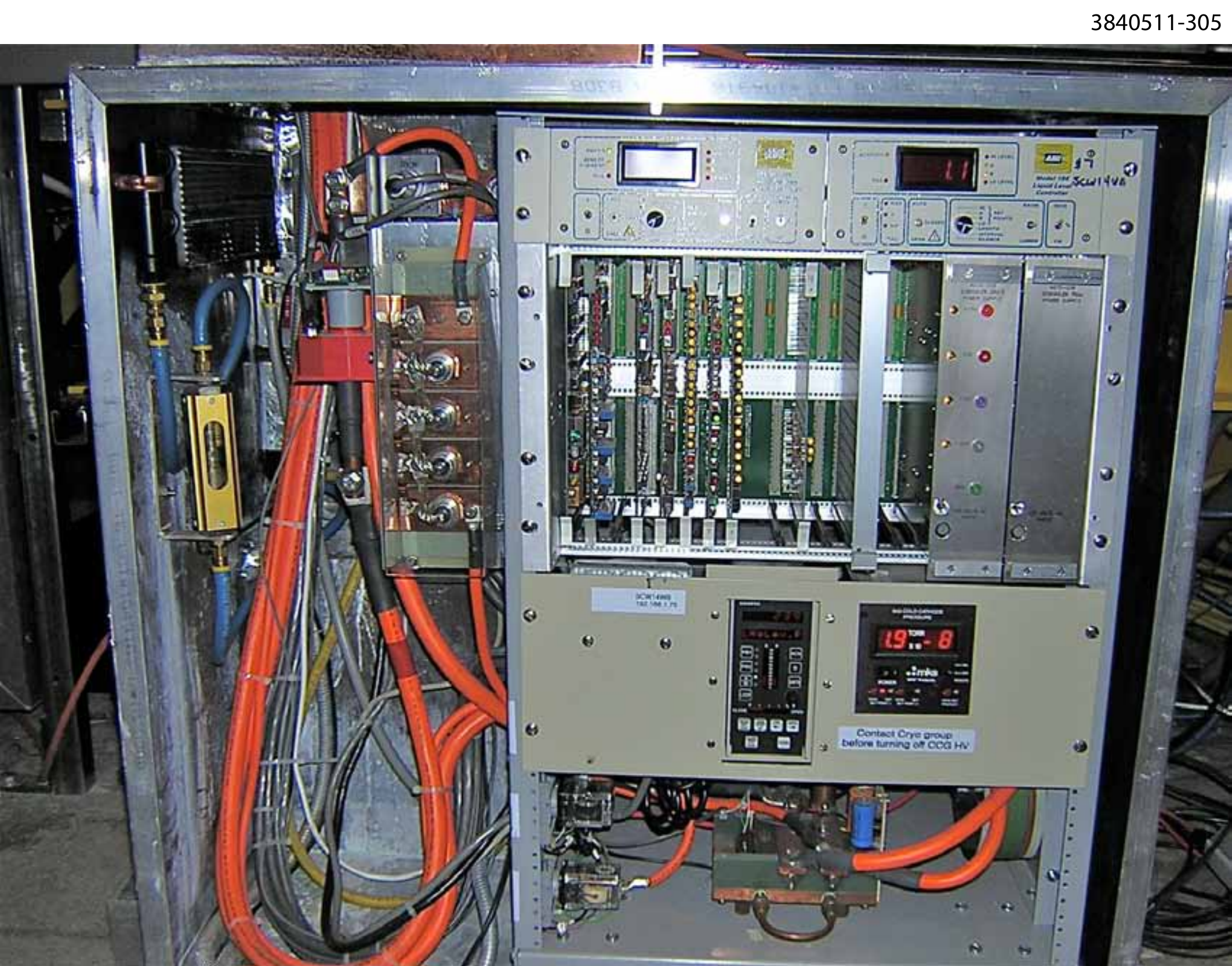}
   \caption[Wiggler controls in the wiggler straight section.]{\label{fig:CtaWigglerControls1}
   This picture presents a view of the wiggler control rack, which contains all of the controls, monitoring and the power supplies for the wiggler and its steering trim coils.}
\end{figure}

In addition to the aforementioned magnet supplies several other types of magnets were either moved or installed around the ring for {\cesrta} use.  These include the wigglers that were moved into the wiggler straight section from their location in the storage ring's arcs between L0 and either L1 or L5 straight sections.  Figure~\ref{fig:CtaWigglerControls2} presents a view of some of the wigglers in the wiggler straight section and figure~\ref{fig:CtaWigglerControls1} shows one of the power supply control racks.  The wiggler straight section's construction required the relocation of six wigglers, their cryogenic controls and power supplies.  The rack of control electronics includes a number of control and monitoring functions.  The first is the ``Ready Chain'' for the protection of the magnets and power supplies, which requires elements such as the primary power, water and cryogenic cooling, the quench protection be enabled and heat sink temperature monitor be below their trip levels, in order for the power supplies to turn on.  The power supplies are a 300~A-3.3~V wiggler main supply and an 8~A-15~V~steering trim supply.  There is also monitoring circuitry including 16~channels of cryogenic temperature readouts, 20~channels of voltage, current, cryogen level and pressure sensor slow readouts, 8~channels of fast quench protection readouts.

The chicane magnets and their power supplies in the L3~experimental area are on loan from SLAC during the {\cesrta} experimental run.  The chicane consists of four dipole magnets wired with field polarities of +, -, -, and +.  They are powered with two DC supplies having a total of 235~A at 56~V.  The system is current regulated using a Danfysic current transformer connected to a modified CESR 16-bit unipolar regulator card, which provides a 10~V maximum control signal to the main DC power supplies.  The integrator in the lead-lag circuit of the regulation loop is adjusted to compensate the magnets' time constant of 0.074~sec.  The modified CESR 16~bit magnet controller card was chosen as it has a common connection to the CESR magnet clock, which allows for coordinated excitation of the chicane magnets with other CESR magnets.  This permits the simultaneous scaling of the current command to the chicane magnets with any other CESR magnet, e.g. to have coordinated scaling of the chicane field with other CESR steering elements.

\subsection{Alignment and Survey Upgrades}
\label{ssec:cesr_conversion.gen_mod.align}

Meeting the more stringent alignment tolerances specified by the {\cesrta} project required an upgrade to both the survey monument hardware and the survey equipment.  All upgrade changes were essentially determined by the conversion of main surveying instrument from a Leica TDM5005 total station to an API Tracker III laser tracker with interferometer.  The laser tracker requires monument hardware, which would accept the 1.5 inch spherically mounted reflectors (SMR's), and also supporting instruments, which could accurately measure the gravity based heights of the reference target.  These were necessary, since at the time of purchase no laser tracker had an integrated high precision gravity based level and compensator.  Thus, a Leica DNA03 digital level and its measurement staffs completed the survey instrument upgrade.

After some study of the available hardware, and a training and observation trip to Lawrence Berkeley National Laboratory's Advanced Light Source we chose to use Hubb's Machine and Manufacturing, Inc. drift nests and floor targets.  Tack welding the drift nests onto Uni-Strut gussets resulted in cost effective and very stable wall targets, which attached to our existing Uni-Strut ribs in the tunnel walls.  The floor targets were permanently epoxied into holes drilled in the concrete floor of the tunnel with a core drill.  The o-rings covers and stainless steel construction of the floor targets provided good protection against water infiltration and corrosion.  Triplets of two wall targets (on inside and outside tunnel walls) and one floor target (near the center of the tunnel floor) spaced at approximately 8 to 10 meters apart provided the necessary geometry to meet the alignment tolerances.

With the utilization of the new survey instruments significant improvements were made in the positioning of accelerator components.  The initial surveys yielded RMS spreads of 206~$\mu$rad for the dipole tilts, 227~$\mu$rad for the quadrupole tilts and 193~$\mu$m for the quadrupole vertical offsets.  After survey and re-alignment the RMS spreads were reduced to 130~$\mu$rad for the dipole tilts, 50~$\mu$rad for the quadrupole tilts and 27~$\mu$m for the quadrupole vertical offsets.\cite{LER2011:Shanks}  With these improvements in the positioning of the accelerator components vertical emittances of 10~pm were achieved.\cite{PRSTAB17:044003}






\section[Beam Instrumentation, Feedback Systems and Injection Controls]{Overview of Beam Instrumentation, Feedback Systems and Injection Controls}
\label{sec:cesr_conversion.beam_instr}

\subsection{Beam Instrumentation, Feedback System Upgrades}
\label{ssec:cesr_conversion.beam_instr.controls}

The undertaking of the \cesrta\ project required the development or upgrading of several accelerator operating systems.  The significant improvements to beam instrumentation are summarized as follows:

\begin{itemize}
\item A major upgrade to the beam position monitor system, which replaced an older relay-based position monitor system with individual readout modules for each monitor capable of turn-by-turn and bunch-by-bunch trajectory measurements for bunches spaced as closely as 4~ns \cite{IPAC10:MOPE089}.
\item The installation of positron and electron x-ray vertical beam size monitors designed for turn-by-turn and bunch-by-bunch beam size measurements for 4~ns spaced bunches \cite{NIMA703:80to90}.
\item Implementation of positron and electron visible-light monitors to measure the horizontal beam size, including the addition of optical elements to allow streak camera measurements of both electron and positron bunches \cite{NIMA767:467to474, NIMA748:96to125}.
\item Development of software to extract bunch-by-bunch tunes utilizing the new modules for the beam position monitors and a second method, which employed video gating of signal from a few beam position monitors from the older relay system \cite{ECLOUD10:PST07, IPAC10:MOPE091}.
\item An upgrade for the tune tracker, which is phase locked to the betatron tunes of a bunch.  This device allows the measurement of the betatron phase advance and the horizontal-to-vertical coupling of CESR permitting their correction \cite{PAC11:MOP215}.
\item Installation of a new beam-stabilizing feedback system, which damps 4~ns-spaced bunches for horizontal, vertical and longitudinal motion, in addition to the existing 14 ns feedback system \cite{PAC01:TPAH006}.
\end{itemize}

This instrumentation is described in further detail in companion papers.  Additionally hardware and software infrastructure was created for X-ray beam size monitors, the CESR beam position monitors and electron cloud detectors.  This included the installation of modest bandwidth and wide bandwidth interfaces to all of these detectors.  It also included the software interfaces and data structures to operate this hardware and to communicate their data to the data acquisition software.

\subsection{Injection Control Upgrades}
\label{ssec:cesr_conversion.beam_instr.injection}

During most of its operations, CESR stored beams with bunches spaced in multiples of 14~ns.  As a result, all of the current monitoring instrumentation was configured for 14~ns as the minimum spacing.  To accommodate the multiple of 4~ns-spaced bunches for {\cesrta}, a new current monitor was developed.  This current monitor utilizes a standard BPM button position monitor connected to one of the new CESR beam position monitor modules, which was programmed specifically to return a constant times the sum of the button signals as the bunch current.  This monitor is capable of measuring all stored bunches with a minimum of a 4~ns-spacing over the desired bunch current range with an update rate of 3~Hz.

The current monitor is an integral component utilized by the injection software.  This program reads the current monitor to obtain the charge in each of the bunches, it then turns on the appropriate triggers to the gun pulser in order to accelerate a set of bunches to fill the storage ring.  Substantial revisions were needed for this software to accommodate more general timing patterns for the stored bunches in CESR.  This is the case since the lowest common harmonic frequency shared by the LINAC injector RF, the LINAC RF accelerator cavities, the synchrotron accelerator cavities and CESR's RF system is 71.4~MHz, equaling a bunch spacing of 14~ns.  Injecting any multiple of 14~ns-spaced bunches only requires turning on triggers for the gun pulser at the correct time and the bunches can be accelerated and be stored in CESR with no change of any RF system phase.  However, injecting 4~ns-spaced bunches requires an additional shift of all of the injector's RF~phases for bunches to line up with proper CESR RF buckets.  The upgraded injection software employed a new operator interface to specify which RF buckets were to be filled in CESR while accomplishing the required triggering for the gun pulser and injector RF phase shifts to store these bunches.  

The injection software had new protection algorithms added, which limits the total current that can be stored in CESR.  This is necessary for several reasons.  When the superconducting wigglers are powered, software interlocks limit the total beam current to prevent their X-ray flux from damaging the downstream vacuum chamber walls.  When the xBSM's optics chip is inserted in the X-ray beam, both software and hardware interlocks limit the total stored beam current.  Also as a failsafe mechanism for communications failure of the bunch-by-bunch current monitor, the injection triggers are disabled whenever the sum of the bunch currents significantly disagrees with the DC current monitor.

\section{Summary}\label{sec:summary}

Major changes were required to modify the CESR storage ring to support the creation of {\cesrta}, a test accelerator configured to study accelerator physics issues over a wide spectrum of accelerator physics
effects and instrumentation related to present light sources and future lepton damping rings.  The modifications were undertaken during four separate accelerator down periods ranging in length from one to four months with these occurring between CESR operating runs for the x-ray science program for CHESS.  When operating for the {\cesrta} program, CESR's optics and instrumentation has been optimized for the study of low emittance tuning methods, electron cloud effects, inter-beam scattering, fast ion instabilities as well as general improvements to beam instrumentation.

\bibliographystyle{amsplain}
\bibliography{Bibliography/CesrTA}







\end{document}